\documentclass[11pt]{article}
\usepackage{moriond,epsfig}
\usepackage{axodraw}

\def\be{\begin{equation}}
\def\ee{\end{equation}}
\def\bea{\begin{eqnarray}}
\def\eea{\end{eqnarray}}

\def\beq{\begin{equation}}
\def\eeq{\end{equation}}
\def\bea{\begin{eqnarray}}
\def\eea{\end{eqnarray}}

\def\Eq#1{Eq.~(\ref{#1})}

\def\ln#1{\mathrm{ln}\left(#1\right)}

\begin{document}
\begin{flushright}
LPN12-074 \\
IFIC/12-47
\end{flushright}
\vspace*{3.5cm}
%\vspace*{4cm}
\title{THE $t\bar t$ ASYMMETRY IN THE STANDARD MODEL AND BEYOND}

\author{Germ\'an Rodrigo}

\address{Instituto de F\'{\i}sica Corpuscular, UVEG - Consejo 
Superior de Investigaciones Cient\'{\i}ficas, \\
Parc Cient\'{\i}fic, 46980 Paterna (Valencia), Spain}

\maketitle\abstracts{
A sizable charge asymmetry in top quark pair 
production has been observed at the Tevatron. 
The experimental results seem to exceed systematically the 
Standard Model theory predictions by a significant amount and have 
triggered a large number of suggestions for 'new physics'. 
The effect is also visible at the LHC, and preliminary 
results have already been presented by the ATLAS and CMS collaborations. 
In this talk, we review the present status of the theoretical 
predictions, and their comparison with the experimental measurements.}

\section{Introduction}

Top quark production at hadron colliders is one of the most active
fields of current theoretical and experimental studies~\cite{Schilling:2012dx},
and one of the most promising probe of physics beyond the Standard Model (SM).
Since 2007, sizable differences have been observed between theory 
predictions~\cite{Antunano:2007da,Kuhn:1998kw,Kuhn:1998jr,Bowen:2005ap} 
for the top quark charge asymmetry and measurements by the CDF~\cite{cdf5,cdf5dilepton,cdf19}
and the D0~\cite{d05,d043,d01} collaborations at the Tevatron. 
This discrepancy was particularly pronounced for the subsample of
$t\bar t$ pairs with large invariant mass, $m_{t\bar t} > 450$~GeV,
and the asymmetry defined in the $t\bar t$ rest-frame, 
where a $3.4 \sigma$ effect was claimed~\cite{cdf5}, although  
recent CDF analysis~\cite{cdf8} lower this discrepancy in the large invariant 
mass region to less than $3 \sigma$. D0 also finds a $3 \sigma$ discrepancy 
when the asymmetry is defined in the leptonic decaying products~\cite{d05}.
It is interesting to note that both experiments find systematically a positive 
excess with respect to the SM.

These discrepancies have triggered a large number of theoretical 
investigations, using these
results, either to restrict new physics like heavy 
axigluons~\cite{Ferrario:2009bz,Rodrigo:2010gm} 
or Kaluza-Klein gluons~\cite{Djouadi:2009nb}
or to postulate a variety of new phenomena in 
the t-channel (u-channel)~\cite{Jung:2009jz,Cheung:2009ch,Shu:2009xf}.
At the same time, the robustness of the leading order QCD prediction 
has been studied in~\cite{Almeida:2008ug,Ahrens:2010zv},
where it has been argued that  next-to-leading (NLL) as well as
next-to-next-to-leading (NNLL) logarithmic corrections do not
significantly modify the leading order result, in agreement with the
approach advocated in~\cite{Kuhn:1998kw,Kuhn:1998jr} 
(Note, however, the large corrections observed  
for the corresponding studies of the $t\bar t$+jet 
sample~\cite{Dittmaier:2008uj}). 

More recently, also the CMS~\cite{CMS,CMS2,CMS3} and ATLAS~\cite{ATLAS}
collaborations have presented the first measurements of the top quark charge 
asymmetry at the LHC. Although the experimental errors are still large, both 
experiments find central values that lie below the SM prediction~\cite{arXiv:1109.6830}, 
in contrast with the Tevatron results. Other measurements at the LHC are 
also heavily constraining the parameter space of the models that have been 
advocated to explain the excess on the charge asymmetry at the Tevatron. 

In this talk we revisit the SM prediction of the top quark charge 
asymmetry at the Tevatron and the LHC~\cite{arXiv:1109.6830}.  
We summarize the experimental measurements of the asymmetry and 
update the pull of their discrepancy with the SM. 
We introduce a new quantity $A_{t\bar t}(Y)$, which measures the 
charge asymmetry with respect to the average rapidity 
of top and antitop quarks, being a suitable observable 
both at the Tevatron and the LHC.
We also analyze the effect of introducing a cut in the $t\bar t$ 
transverse momentum as a possible explanation of the discrepancy.
Finally, we comment on beyond the SM contributions to the asymmetry

\section{The charge asymmetry in the SM}

\begin{figure}[htb]
\begin{center}

\begin{picture}(200,70)(0,0)
\SetWidth{1.1}
\SetScale{.7}

\SetOffset(-10,40)

\Vertex(40,0){2}
\Vertex(70,0){2}
\Vertex(0,30){2}
\Vertex(0,-30){2}
\Vertex(-60,30){2}
\Vertex(-60,-30){2}
\Gluon(-60,30)(0,30){5}{6}
\Gluon(0,-30)(-60,-30){5}{6}
\Gluon(40,0)(70,0){4}{3}
\ArrowLine(0,30)(0,-30)
\ArrowLine(0,-30)(40,0)
\ArrowLine(40,0)(0,30)
\ArrowLine(100,30)(70,0)
\ArrowLine(70,0)(100,-30)
\ArrowLine(-90,30)(-60,30)
\ArrowLine(-60,30)(-60,-30)
\ArrowLine(-60,-30)(-90,-30)
\DashLine(25,30)(25,-30){5}
\DashLine(20,40)(-20,-40){5}
\Text(0,-45)[]{(a)}

\SetOffset(195,40)

\Vertex(40,0){2}
\Vertex(70,0){2}
\Vertex(0,30){2}
\Vertex(0,-30){2}
\Vertex(-60,30){2}
\Vertex(-60,-30){2}
\Gluon(-60,30)(0,-30){5}{8}
\Gluon(0,30)(-60,-30){5}{8}
\Gluon(40,0)(70,0){5}{3}
\ArrowLine(0,30)(0,-30)
\ArrowLine(0,-30)(40,0)
\ArrowLine(40,0)(0,30)
\ArrowLine(100,30)(70,0)
\ArrowLine(70,0)(100,-30)
\ArrowLine(-90,30)(-60,30)
\ArrowLine(-60,30)(-60,-30)
\ArrowLine(-60,-30)(-90,-30)
\DashLine(25,30)(25,-30){5}
\DashLine(20,40)(-20,-40){5}
\Text(0,-45)[]{(b)}

\end{picture}
\end{center}
\caption{Cut diagrams representing the QCD contribution to the charge asymmetry.\label{fig:cut}}
\end{figure}
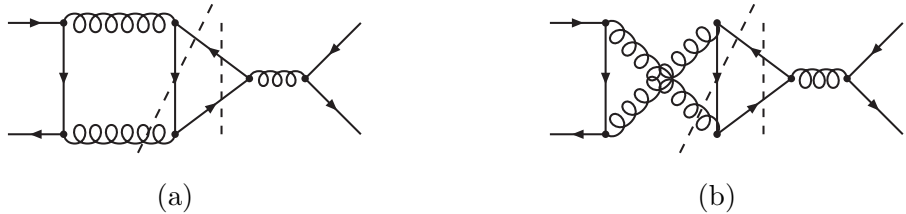

The dominant contribution to the charge asymmetry
originates from $q\bar{q}$ annihilation~\cite{Kuhn:1998kw,Kuhn:1998jr}. 
Specifically, it originates from the interference between the 
Born amplitudes for $q\bar{q}\to Q\bar{Q}$ and the part of the one-loop 
correction, which is antisymmetric under the exchange of the heavy quark 
and antiquark (box and crossed box). 
To compensate the infrared divergences, this virtual correction 
must be combined with the interference between initial and 
final state radiation.
Diagrams with triple gluon coupling in both real and virtual 
corrections give rise to symmetric amplitudes
and can be ignored. The corresponding contribution to the rate is conveniently 
expressed by the absorptive contributions (cuts) of the 
diagrams depicted in Fig~\ref{fig:cut}.
A second contribution to the asymmetry from 
quark-gluon scattering (``flavor excitation'') hardly contributes to 
the asymmetry at the Tevatron. At the LHC, it enhances 
the asymmetry in suitable chosen kinematical regions~\cite{Kuhn:1998jr}. 
CP violation arising from electric or chromoelectric dipole moments 
of the top quark do not contribute to the asymmetry, 
unless the asymmetry is defined through the decay products.

%%%%%%%%%%%%%%%%%%%%%%%%%%%%%%%%%%%%%%%%%%%%%%%
%%%%%%%%%%%%%%%%%%%%%%%%%%%%%%%%%%%%%%%%%%%%%%%
%%%%%%%%%%%%%%%%%%%%%%%%%%%%%%%%%%%%%%%%%%%%%%%
%%%%%%%%%%%%%%%%%%%%%%%%%%%%%%%%%%%%%%%%%%%%%%%

The inclusive charge asymmetry is proportional to the symmetric colour 
factor $d_{abc}^2=40/3$, and it is positive, namely the 
top quarks are preferentially emitted in the direction of the incoming 
quarks at the partonic level~\cite{Kuhn:1998kw,Kuhn:1998jr}. 
The colour factor can be understood from 
the different behaviour under charge conjugation of the scattering amplitudes 
with the top and antitop quark pair in a colour singlet or colour octet state.  
The positivity of the inclusive asymmetry is a consequence of the fact that 
the system will be less perturbed, or in other words will require less energy, 
if the outgoing colour field flows in the same direction as the incoming 
colour field. On the other hand, radiation of gluons requires to decelerate 
the colour charges, and thus the asymmetry of the $t\bar t$+jet sample is negative.   

\begin{figure}[t]
\begin{center}
\includegraphics[width=5.5cm]{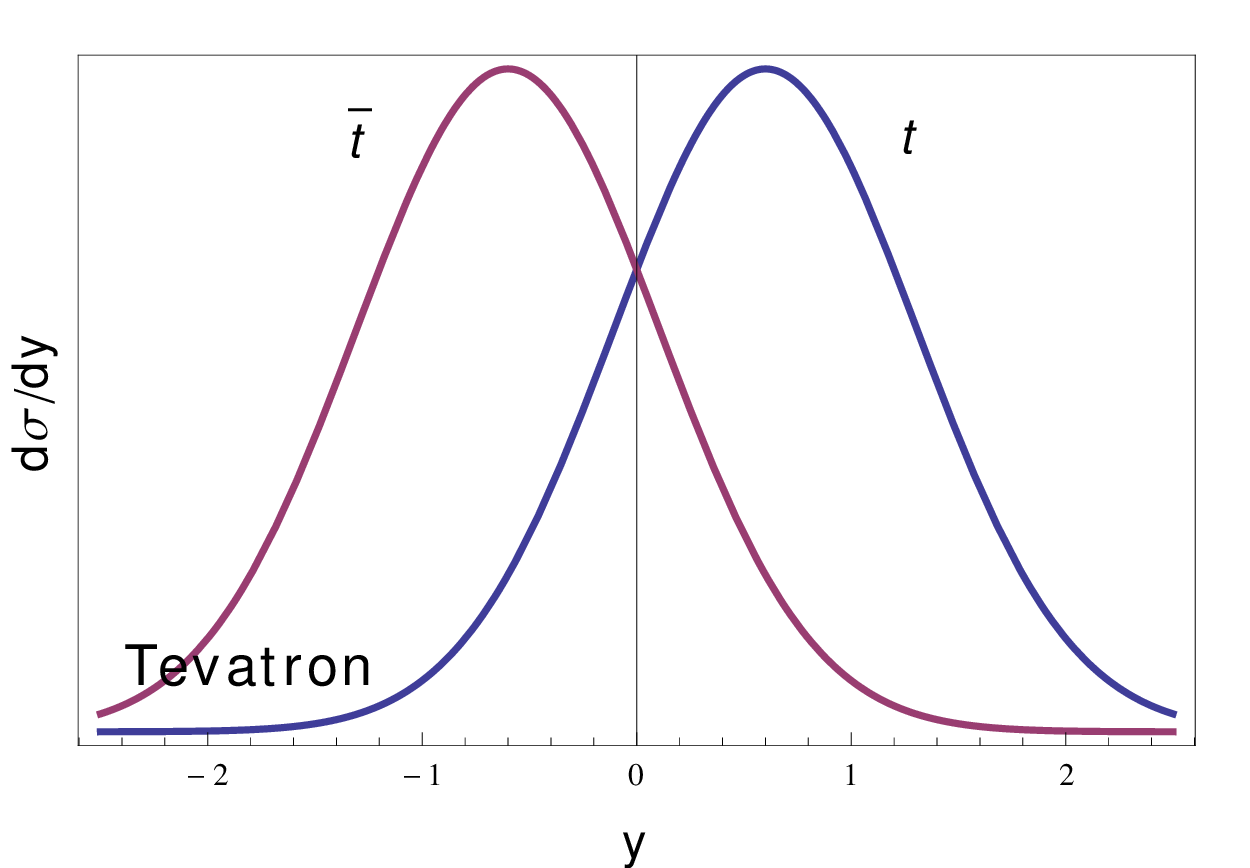} \qquad
\includegraphics[width=5.5cm]{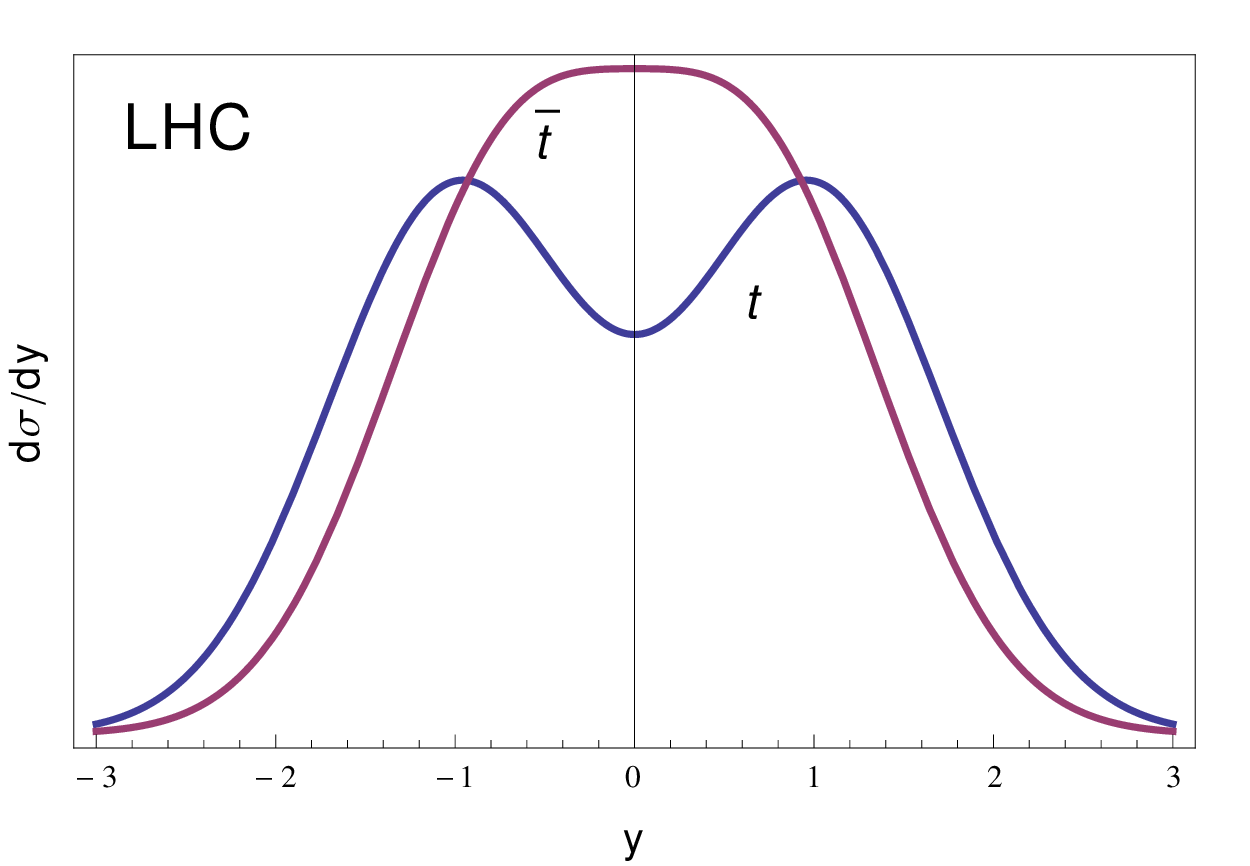}
\caption{Not to scale rapidity distributions of top and antitop quarks 
at the Tevatron (left) and the LHC (right).
\label{fig:rap}}
\end{center}
\end{figure}

At Tevatron, the charge asymmetry is equivalent to a forward--backward
asymmetry as a consequence of charge conjugation symmetry, and thus 
top quarks are preferentially emitted in the direction of the incoming protons. 
The charge asymmetry can also be investigated in proton-proton
collisions at the LHC~\cite{Antunano:2007da,Kuhn:1998kw,Kuhn:1998jr} 
by exploiting the small $t\bar t$ sample produced in annihilation 
of valence quarks and antiquarks from the sea. Since valence quarks carry 
on average more momentum than sea antiquarks,
production of top quarks with larger rapidities will be preferred in the SM, 
and antitop quarks will be produced more frequently at smaller 
rapidities. Figure~\ref{fig:rap} shows for comparison, qualitatively 
and not to scale, the rapidity distributions of the top and the antitop quarks 
at the Tevatron (left) and the LHC (right).

%%%%%%%%%%%%%%%
%%%%%%%%%%%%%%%

Diagrams similar to those depicted in Fig.~\ref{fig:cut}, where
one of the gluons has been substituted by a photon, also lead 
to a contribution to the charge asymmetry from mixed QED-QCD
corrections. The relative factor between QCD and QED 
asymmetries amounts to 
\beq
f_q^{\rm QED} = 3 \, \frac{\alpha_{\rm QED} \, Q_t \, Q_q}
{\displaystyle \frac{\alpha_S}{2} \, \left( \frac{d_{abc}^2}{4}\right)^2}
= \frac{\alpha_{\rm QED}}{\alpha_S} \, \frac{36}{5} \, Q_t \, Q_q
\label{eq:fqQED}
\eeq
for one quark species, and to 
\beq
f^{\rm QED} = \frac{4 f_u^{\rm QED} +  f_d^{\rm QED}}{5} = 
\frac{\alpha_{\rm QED}}{\alpha_S} \, \frac{56}{25} \approx 0.18~,
\label{eq:QED}
\eeq
after convolution with the PDFs if one considers as a first approximation 
that the relative importance of $u\bar u$ versus $d\bar d$ 
annihilation at the Tevatron is $4:1$. 
Thus, to an enhancement of nearly twenty percent 
of the QCD asymmetry, in good agreement with the more detailed 
numerical studies of~\cite{arXiv:1109.6830,Hollik:2011ps}.
At the LHC, the relative importance of $u\bar u$ versus $d\bar d$ 
annihilation is approximately $2:1$, thus reducing $f^{\rm QED}$ 
down to $0.13$. Similarly, weak contributions with the 
photon replaced by the $Z$ boson should be considered at the same 
footing. However, as a consequence of the cancellation between up and 
down quark contributions, and the smallness of the weak coupling, 
the weak corrections at the Tevatron are smaller by more than a factor 
$10$ than the corresponding QED result. 
For proton-proton collisions the cancellation between up and down 
quark contributions is even stronger and the total weak correction 
is completely negligible.

%%%%%%%%%%%%%%%%%%%%%%%%%%%%%%%%%%%%%%%%%%%%%%%%%%%%%%%%%%%%%%%
%%%%%%%%%%%%%%%%%%%%%%%%%%%%%%%%%%%%%%%%%%%%%%%%%%%%%%%%%%%%%%%
%%%%%%%%%%%%%%%%%%%%%%%%%%%%%%%%%%%%%%%%%%%%%%%%%%%%%%%%%%%%%%%

\section{SM predictions of the charge asymmetry at the Tevatron and the LHC}

The charge asymmetry at the Tevatron is equivalent to a 
forward--backward asymmetry. In the laboratory frame it is given 
by either of the following definitions
\beq
A_{\rm lab}=\frac{N(y_t>0)-N(y_t<0)}{N(y_t>0)+N(y_t<0)}
= \frac{N(y_t>0)-N(y_{\bar t}>0)}{N(y_t>0)+N(y_{\bar t}>0)}~, 
\eeq
requiring to measure the rapidity of either the $t$ or 
the $\bar t$ for each event. 
The most recent experimental analysis measure both 
rapidities simultaneously, and define the asymmetry 
in the variable $\Delta y = y_t-y_{\bar t}$, which is invariant 
under boosts, and thus equivalent to measure the charge asymmetry in 
the $t \bar t$ rest-frame:
\beq
A_{t\bar t}=\frac{N(\Delta y > 0)-N(\Delta y < 0)}
{N(\Delta y > 0)+N(\Delta y < 0)}~.
\label{Attbar}
\eeq
The size of the charge asymmetry in the 
$t\bar t$ rest-frame is about $50\%$ larger than in the laboratory 
frame~\cite{Antunano:2007da} because part of the asymmetry 
is washed out by the boost from the partonic rest-frame to 
the laboratory. 

At the LHC, the charge asymmetry has been defined~\cite{CMS2,ATLAS}
through $\Delta |y| = |y_t|-|y_{\bar t}|$~\footnote{CMS~\cite{CMS2} has also used
pseudorapidities to define the charge asymmetry with 
$\Delta |\eta| = |\eta_t|-|\eta_{\bar t}|$. The size of the asymmetry in $\Delta |\eta|$
is only slightly higher~\cite{arXiv:1109.6830} than with $\Delta |y|$.}
\beq
A_C^y = \frac{N(\Delta |y|>0)-N(\Delta |y|<0)}
{N(\Delta |y|>0)+N(\Delta |y|<0)}~.
\label{AC}
\eeq 
A forward--backward asymmetry obviously vanishes in a symmetric machine like the LHC. 

%%%%%%%%%%%%%%%
\begin{figure}[t]
\begin{center}
\includegraphics[width=6.3cm]{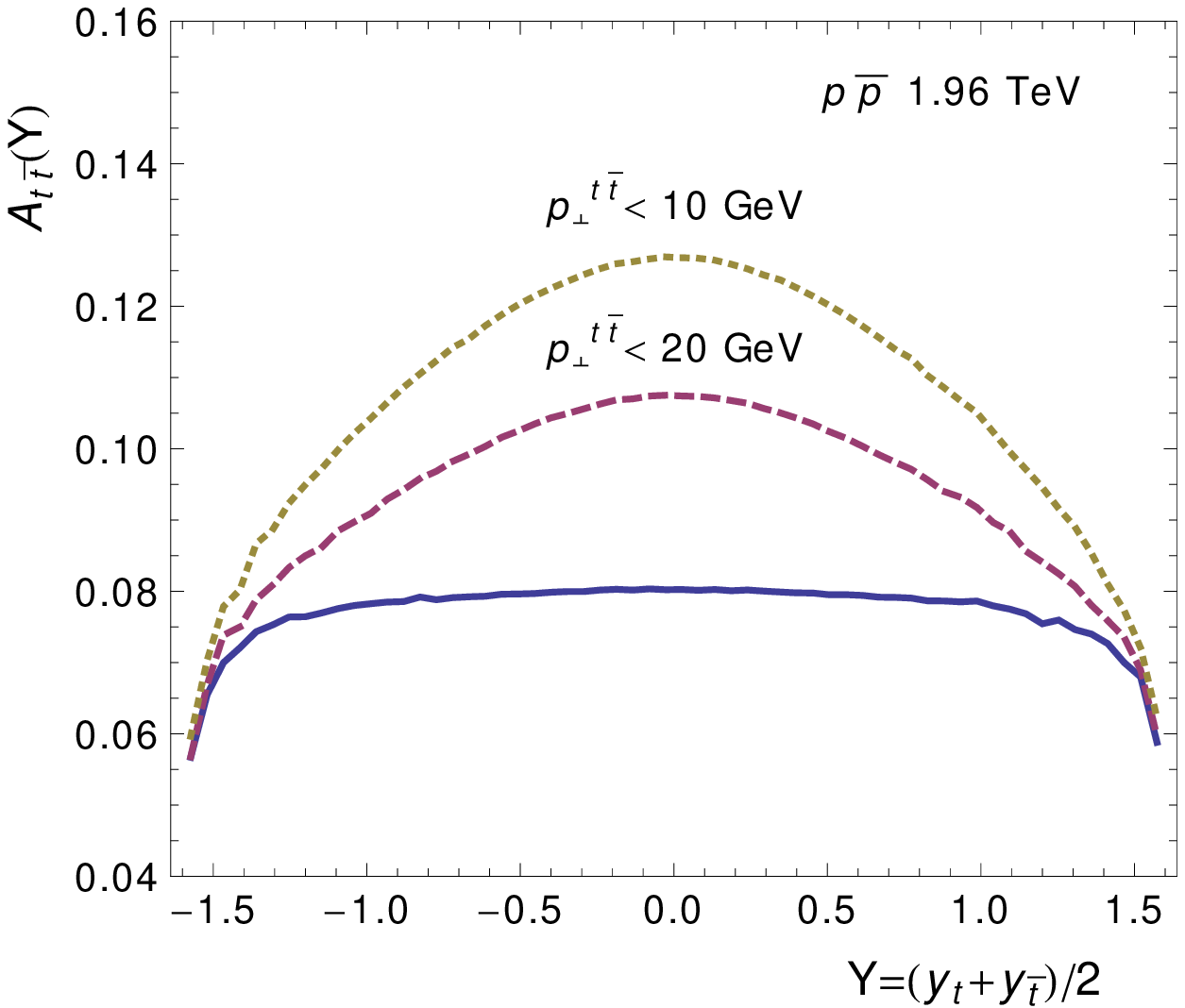} \qquad
\includegraphics[width=6.5cm]{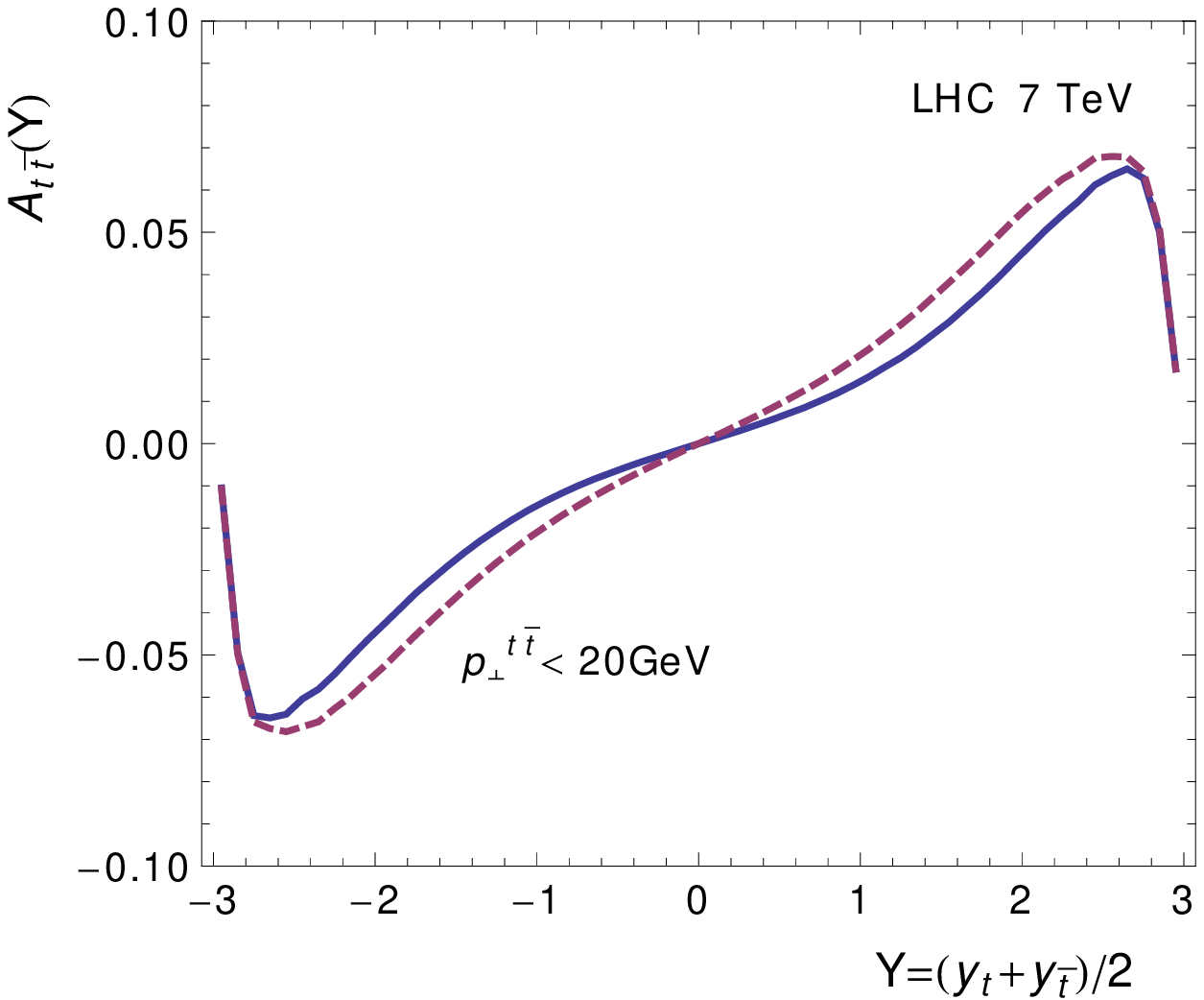}
\caption{Universal charge asymmetry $A_{t\bar t} (Y)$ as a function of  
the mean rapidity $Y=(y_t+y_{\bar t})/2$. 
Solid line: without cut on $p_\perp^{t\bar t}$, 
dotted/dashed lines: with cut on $p_\perp^{t\bar t}$.
\label{fig:AY}}
\end{center}
\end{figure}
%%%%%%%%%%%%%%%

The $t\bar t$ asymmetry is thus often called forward--backward asymmetry 
at the Tevatron and charge asymmetry at the LHC, but in fact, although the 
kinematical configurations of the two machines are different the physical 
origin of the asymmetry in both cases is the same (see Fig.~\ref{fig:cut}). 
However, it is possible to define a universal observable, namely an asymmetry 
suitable for both the Tevatron and the LHC, if we measure the charge asymmetry 
with respect to the average rapidity $Y=(y_t+y_{\bar t})/2$ of the top and the 
antitop quarks. This universal charge asymmetry~\cite{arXiv:1109.6830} is 
obtained by selecting events for a definite average rapidity $Y$ and 
calculating their asymmetry as in \Eq{Attbar}:
\beq
A_{t\bar t}\, (Y)=\frac{N(\Delta y > 0)-N(\Delta y < 0)}
{N(\Delta y > 0)+N(\Delta y < 0)}~.
\label{eq:pair}
\eeq
The theoretical prediction for the differential distribution $A_{t\bar t} (Y)$ 
as a function of $Y$ is shown in Fig.~\ref{fig:AY}~(left) for the Tevatron, and 
in Fig.~\ref{fig:AY}~(right) for the LHC.  
By construction $A_{t\bar t} (Y)$ is a symmetric function of $Y$ at 
the Tevatron, and an antisymmetric function of $Y$ at the LHC.
At the Tevatron the asymmetry $A_{t\bar t} (Y)$ is almost flat, 
at the LHC it resembles a forward--backward asymmetry. The corresponding 
integrated asymmetries coincide with the usual charge asymmetry in 
the $t\bar t$ rest-frame from \Eq{Attbar}, $A_{t\bar t} (Y) \to A_{t\bar t}$,
and with the charge asymmetry $A_C^y$ in \Eq{AC} if we select events 
with $Y$ either positive or negative. The advantage of \Eq{eq:pair}
for the LHC is that the size of the asymmetry can be enhanced by 
selecting events with a minimum average rapidity $Y>Y_{\rm cut}$~\cite{arXiv:1109.6830}.
This is relevant because $t\bar t$ production at the LHC, contrary to 
what happens at the Tevatron, is dominated by gluon fusion 
which is symmetric. Therefore, in order to reach a sizable asymmetry at the LHC
it is necessary to introduce selection cuts to suppress as much as possible 
the contribution of gluon fusion events, and to enrich the sample 
with $q\bar q$ events. In particular, gluon fusion is dominant in the central 
region and can be suppressed by introducing a cut in the average rapidity $Y$
(or selecting events with large $m_{t\bar t}$).
Obviously this is done at the price of lowering the statistics, which, 
however, is not a problem at the LHC. 
  
%%%%%%%%%%%%%%%
\begin{table}[t]
\begin{center}
\caption{SM asymmetries in the laboratory $A_{\rm lab}$
and the $t\bar t$ rest-frame $A_{t \bar t}$ at Tevatron. 
Predictions are given also for samples with the top quark 
pair invariant mass $m_{t\bar t}$ above and below $450$~GeV, 
and with $|\Delta y|=|y_t-y_{\bar t}|$ larger or smaller than one.
Summary of latest experimental results: numbers with $^*$ refer 
to "reconstruction level'', the others to parton level. 
The former cannot be compared directly with the quoted theoretical 
predictions. 
\label{tab:Attbar}}
\begin{tabular}{|c|ccccc|} \hline

Tevatron & inclusive      & $m_{t\bar t}< 450$ GeV & $m_{t\bar t}> 450$ GeV 
            & $|\Delta y|<1$ & $|\Delta y|>1$   \\ \hline 
SM laboratory  $A_{\rm lab}$ & 0.056 (7) & 0.029 (2) & 0.102 (9) & & \\ \hline 
CDF~\cite{cdf5} & 0.150 (55) & 0.059 (34)$^*$ & 0.103 (49)$^*$ &  &  \\ \hline \hline 
SM $t\bar t$ rest-frame $A_{t \bar t}$  & 0.087 (10)& 0.062 (4) & 0.128 (11) 
                & 0.057 (4) & 0.193 (15) \\ \hline
D0~\cite{d05}   & 0.196 (65) & 0.078 (48)$^*$ & 0.115 (60)$^*$ & 0.061 (41)$^*$ & 0.213 (97)$^*$ \\ 
CDF~\cite{cdf8} & 0.162 (47) & 0.078 (54) & 0.296 (67) & 0.088 (47) & 0.433 (109) \\ \hline
\end{tabular}

\end{center}
\end{table}
%%%%%%%%%%%%%%%

%%%%%%%%%%%%%%%
\begin{figure}[t]
\begin{center}
\includegraphics[width=8cm]{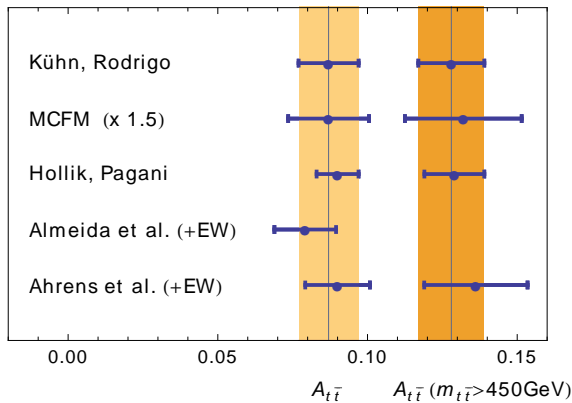}
\caption{Summary of theoretical predictions for the inclusive charge asymmetry 
at the Tevatron in the $t\bar t$ rest-frame, $A_{t\bar t}$, and in the large 
invariant mass region $A_{t\bar t}(m_{t\bar t}>450 {\rm GeV})$.
\label{fig:th}}
\end{center}
\end{figure}
%%%%%%%%%%%%%%%

%Let us summarize the state-of-the-art SM predictions
%for the charge asymmetry. 
Predictions in the SM for the charge asymmetry at the Tevatron 
in the laboratory frame and in the $t\bar t$ rest-frame  
are listed~\cite{arXiv:1109.6830} in Table~\ref{tab:Attbar}.
In order to compare theoretical results in the SM with 
the most recent measurements at Tevatron,
predictions in Table~\ref{tab:Attbar} are presented also for samples
with $m_{t\bar t}$ larger and smaller than $450$~GeV, and 
with $|\Delta y| = |y_t-y_{\bar t}|$ larger and smaller than one.
These predictions include also the QED and weak (strongly suppressed) 
corrections. Those corrections enhance the QCD asymmetry by  
an overall factor $1.21$, which is slightly different 
from \Eq{eq:QED} due to the deviation of the relative amount of 
$u\bar u$ and $d\bar d$ contributions from the simple approximation $4:1$. 

The charge asymmetry is the ratio of the antisymmetric cross-section 
to the symmetric cross-section. The leading order contribution to the 
antisymmetric cross-section is a loop effect (Fig.~\ref{fig:cut}), 
but the leading order contribution to the symmetric cross-section 
appears at the tree-level. This suggest that the charge asymmetry 
should be normalized to the Born cross-section~\cite{Kuhn:1998kw,Kuhn:1998jr}, 
and not the NLO cross-section~\cite{cdf5}, in spite of the fact that 
the later is well known, and is included in several Monte Carlo event 
generators such as MCFM~\cite{Campbell:1999ah}.
This procedure~\cite{Kuhn:1998kw,Kuhn:1998jr} is furthermore supported 
by the fact that theoretical predictions resuming leading logarithms 
(NLL~\cite{Almeida:2008ug} and NNLL~\cite{Ahrens:2010zv})
do not modify significantly the central prediction for the asymmetry. 

Figure~\ref{fig:th} summarizes the state-of-the-art SM predictions 
for the inclusive asymmetry in the $t\bar t$ rest-frame, and 
in the large invariant mass region, $m_{t\bar t}>450$~GeV, from different 
authors~\cite{arXiv:1109.6830,cdf5,Hollik:2011ps,Almeida:2008ug,Ahrens:2010zv}. 
In order to have a coherent picture, EW corrections have been added
to the predictions presented in~\cite{cdf5,Almeida:2008ug,Ahrens:2010zv}, 
which amount to a factor of about $1.2$, and the Monte Carlo based 
prediction has also been corrected by an extra factor of $1.3$ 
to account for the normalization to the NLO cross-section.
A nice agreement if found among the different theoretical predictions.  

There is, moreover, an intense effort in the community to evaluate the 
$t\bar t$ cross-section at the next-to-next-to-leading order
(NNLO)~\cite{Bonciani:2010ue,Moch:2012mk}. 
First results have been obtained recently for the channel 
$q\bar q \to t \bar t$~\cite{Baernreuther:2012ws}.
Thus, all the relevant ingredients to calculate the asymmetry 
at the next order are available; NNLO corrections to the gluon fusion 
channel are not necessary if the asymmetry is normalized to the 
NLO cross-section.

%%%%%%%%%%%%%%%%%%%%%%%%%%%%%%%%%%%%%%%%%%%%%%%%%%%%%%%%%%%%%

\begin{table}[t]
\begin{center}
\caption{SM charge asymmetries $A_C^y$,
and integrated universal charge asymmetry $A_{t\bar t}^{cut}(Y_{\rm cut}=0.7)$, 
at different LHC energies. Summary of recent measurements by CMS and ATLAS. 
\label{tab:AttbarmultiTeV}}
\vspace{0.4cm}
\begin{tabular}{|l|cc|} \hline     
               &  $A_C^y$      &  $A_{t\bar t}^{\rm cut}(Y_{\rm cut}=0.7)$  \\ \hline 
LHC 7 TeV      &  0.0115 (6) &  0.0203 (8)  \\
LHC 8 TeV      &  0.0102 (5) &  0.0178 (6)  \\
LHC 14 TeV     &  0.0059 (3) &  0.0100 (4)  \\\hline\hline
LHC 7 TeV CMS~\cite{CMS}     & 0.004 $\pm$ 0.010 $\pm$ 0.012  & \\
LHC 7 TeV ATLAS~\cite{ATLAS} & -0.018 $\pm$ 0.028 $\pm$ 0.023   & \\\hline
\end{tabular}
\end{center}
\end{table}

%%%%%%%%%%%%%%%%%%%%%%%%%%%%%%%%%%%%%%%%%%%%%%%%%%%%%%%%%%%%%

%%%%%%%%%%%%%%%
\begin{figure}[t]
\begin{center}
\includegraphics[width=8cm]{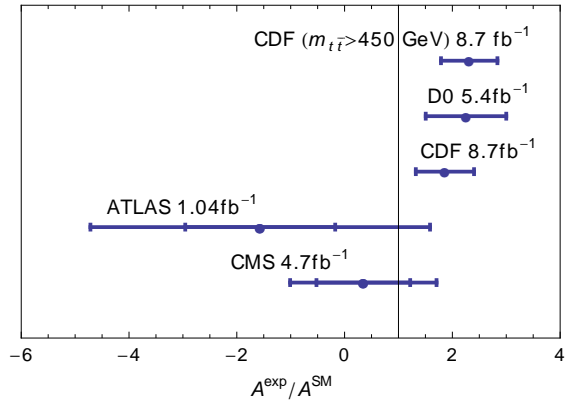}
\caption{Comparison of some of the most recent measurements of the charge 
asymmetry at the Tevatron and the LHC with the corresponding SM predictions.
\label{fig:tension}}
\end{center}
\end{figure}
%%%%%%%%%%%%%%%

The SM predictions for the charge asymmetry $A_C^y$ in \Eq{AC} are 
listed in Table~\ref{tab:AttbarmultiTeV} for different center-of-mass 
energies of the LHC, together with the most recent experimental 
measurements at $\sqrt{s}=7$~TeV. It is interesting to note that 
both experiments obtain central values for the asymmetry 
that are below the SM prediction. These results, although compatible with 
the SM prediction within uncertainties, are in some "tension'' with 
the Tevatron measurements. Unless different selection cuts are introduced, 
the signs of the asymmetry at the Tevatron and the LHC are generally correlated.
A quantitative estimation of this "tension'' is shown in Fig.~\ref{fig:tension}. 
It amounts to about 1$\sigma$ or below, and thus it is still non conclusive. 
New analysis with larger statistics should be expected soon, and will 
reduce further the experimental errors. In fact, given the amount of data 
expected to be collected in the current 2012 run of the LHC, 
the measurements of the asymmetry will become soon dominated by systematics,
and not by statistics as for the Tevatron.

Unfortunately, the asymmetry at the LHC decreases at higher energies because 
of the larger gluon fusion contribution. It can, however,  
be enhanced by selecting events with large rapidities or large $m_{t\bar t}$.
Theoretical predictions for the universal charge asymmetry in \Eq{eq:pair} 
with $Y > 0.7$ are also presented in Table~\ref{tab:AttbarmultiTeV}.

\section{Explaining the discrepancy with the SM}

In the last years, hundreds of papers have postulated new physics models 
to explain the discrepancy with the SM, particularly after the publication 
of the CDF measurement in the high invariant mass region~\cite{cdf5}.
A new CDF measurements with larger statistics~\cite{cdf8} define better 
the slope of the $m_{t\bar t}$ distribution, showing a persistent depart from 
the SM, although slightly reducing the discrepancy. 
Although smaller than measured, the SM definitely predicts a positive slope 
in the $m_{t\bar t}$ distribution.  This effect can 
be understood from the fact that events with real emission of gluons give a negative 
contribution to the asymmetry, and from the fact that $m_{t\bar t}< \sqrt{\hat s}$, 
where $\hat s$ is the partonic center-of-mass energy of each event. As a consequence,
negative contributions to the asymmetry prefer to be at low values of $m_{t\bar t}$, 
while the high invariant mass region receives less bremsstrahlung contributions.
D0~\cite{d05} also shows results which are consistent with CDF. Thus, 
CDF and D0 analysis with full statistics or the combination of both experiments 
will provide more accurate results in the future but it is unlikely that the 
new measurements will differ significantly with previous results. 

There has been a recent discussion about the distribution of the transverse momentum 
of the $t\bar t$ pair~\cite{arXiv:1109.6830,d05,cdf8,Skands:2012mm}, $p_{\perp}^{t\bar t}$. 
An inaccurate simulation of the $p_{\perp}^{t\bar t}$ distribution could lead to a 
mismatch in the estimate of the negative contributions to the asymmetry, and 
thus to an asymmetry much larger than expected. 
On the other hand, it has also been suggested that the $p_{\perp}^{t\bar t}$ distribution 
could be used to enhance the size of the asymmetry~\cite{Alvarez:2012uh}. In that 
case theoretical prediction might be affected by large $\ln{p_{\perp}^{t\bar t}}$ that 
need to be resumed. This issues should be further investigated.

Since the asymmetry is proportional to the strong coupling $\alpha_S(\mu)$ 
a larger asymmetry can be obtained by conveniently choosing a very 
small value of the renormalization scale~\cite{Brodsky:2012ik}. 
However, a fine tuning of $\alpha_S(\mu)$ does not guarantee the 
convergence of the perturbative series at higher orders, 
and would bring the LHC results in disagreement with the SM. 

It is not the purpose of this talk to make a complete review of 
new physics models explaining the Tevatron anomaly on the charge 
asymmetry. It is, however, interesting to mention that new 
LHC results, not only on direct searches in the dijet 
or $t\bar t$ differential cross-sections, but also in same-sign 
top quark production~\cite{Aad:2012bb}, or 
$t \bar t$+jet~\cite{Chatrchyan:2012su}, are seriously constraining
the parameter space of these models.

\section{Summary}

Tevatron has shown in the last years a systematic upward discrepancy 
in the measurement of the top quark charge asymmetry with respect 
to theoretical predictions in the SM. These discrepancies have 
triggered a large number of theoretical speculations about 
possible contributions beyond the SM. The LHC experiments have 
also presented the first measurements of the charge asymmetry, 
and due to his present good performance, will be able to provide 
much more accurate and competitive measurements after the 2012 run. 
New theoretical developments should also help to shed light on 
the Tevatron anomaly. Certainly, 2012 will be a crucial date to solve 
this puzzle.

\section*{Acknowledgments}

Supported by REA Grant Agreement PITN-GA-2010-264564 (LHCPhenoNet), 
by the MICINN Grant No. FPA2007-60323, and FPA2011-23778, 
by CPAN (Grant No. CSD2007-00042), and
by the Generalitat Valenciana Grant No. PROMETEO/2008/069.
Thanks to J. H. K\"uhn for a very fruitful collaboration.

\end{document}